\documentstyle[12pt]{article}

  \def\pp{{\mathchoice
          {
              \kern 1pt%
              \raise 1pt
              \vbox{\hrule width5pt height0.4pt depth0pt
                    \kern -2pt
                    \hbox{\kern 2.3pt
                          \vrule width0.4pt height6pt depth0pt
                          }
                    \kern -2pt
                    \hrule width5pt height0.4pt depth0pt}%
                    \kern 1pt
           }
            {
              \kern 1pt%
              \raise 1pt
              \vbox{\hrule width4.3pt height0.4pt depth0pt
                    \kern -1.8pt
                    \hbox{\kern 1.95pt
                          \vrule width0.4pt height5.4pt depth0pt
                          }
                    \kern -1.8pt
                    \hrule width4.3pt height0.4pt depth0pt}%
                    \kern 1pt
            }
            {
              \kern 0.5pt%
              \raise 1pt
              \vbox{\hrule width4.0pt height0.3pt depth0pt
                    \kern -1.9pt  
                    \hbox{\kern 1.85pt
                          \vrule width0.3pt height5.7pt depth0pt
                          }
                    \kern -1.9pt
                    \hrule width4.0pt height0.3pt depth0pt}%
                    \kern 0.5pt
            }
            {
              \kern 0.5pt%
              \raise 1pt
              \vbox{\hrule width3.6pt height0.3pt depth0pt
                    \kern -1.5pt
                    \hbox{\kern 1.65pt
                          \vrule width0.3pt height4.5pt depth0pt
                          }
                    \kern -1.5pt
                    \hrule width3.6pt height0.3pt depth0pt}%
                    \kern 0.5pt
            }
        }}

  \def\mm{{\mathchoice
                       {
                             \kern 1pt
               \raise 1pt    \vbox{\hrule width5pt height0.4pt depth0pt
                                  \kern 2pt
                                  \hrule width5pt height0.4pt depth0pt}
                             \kern 1pt}
                       {
                            \kern 1pt
               \raise 1pt \vbox{\hrule width4.3pt height0.4pt depth0pt
                                  \kern 1.8pt
                                  \hrule width4.3pt height0.4pt depth0pt}
                             \kern 1pt}
                       {
                            \kern 0.5pt
               \raise 1pt
                            \vbox{\hrule width4.0pt height0.3pt depth0pt
                                  \kern 1.9pt
                                  \hrule width4.0pt height0.3pt depth0pt}
                            \kern 1pt}
                       {
                           \kern 0.5pt
             \raise 1pt  \vbox{\hrule width3.6pt height0.3pt depth0pt
                                  \kern 1.5pt
                                  \hrule width3.6pt height0.3pt depth0pt}
                           \kern 0.5pt}
                       }}

\catcode`@=11
\def\un#1{\relax\ifmmode\@@underline#1\else
        $\@@underline{\hbox{#1}}$\relax\fi}
\catcode`@=12

\let\du=\du                     

\def\a{\alpha}
\def\b{\beta}

\def\d{\delta}

\def\f{\phi}

\def\j{\psi}

\def\l{\lambda}
\def\m{\mu}

\def\q{\theta}

\def\s{\sigma}

\def\G{\Gamma}

\def\L{\Lambda}
\def\O{\Omega}
\def\P{\Pi}

\def\S{\Sigma}
\def\U{\Upsilon}
\def\X{\Xi}

\def\ve{\varepsilon}

\def\cc{{\cal C}}

\def\ce{{\cal E}}

\def\cm{{\cal M}}

\def\car{{\cal R}}

\def\cx{{\cal X}}
\def\cy{{\cal Y}}

\def\bo{{\raise-.5ex\hbox{\large$\Box$}}}               
\def\pa{\partial}                                       
\def\de{\nabla}                                         
\def\TH{{\raise.2ex\hbox{$\displaystyle \bigodot$}\mskip-4.7mu \llap H \;}}
\def\face{{\raise.2ex\hbox{$\displaystyle \bigodot$}\mskip-2.2mu \llap {$\ddot
        \smile$}}}                                      

\def\sp#1{{}^{#1}}                              
\def\Bar#1{\overline{#1}}                       
\def\leftrightarrowfill{$\mathsurround=0pt \mathord\leftarrow \mkern-6mu
        \cleaders\hbox{$\mkern-2mu \mathord- \mkern-2mu$}\hfill
        \mkern-6mu \mathord\rightarrow$}
\def\dvec#1{\vbox{\ialign{##\crcr
        \leftrightarrowfill\crcr\noalign{\kern-1pt\nointerlineskip}
        $\hfil\displaystyle{#1}\hfil$\crcr}}}           

\def\frac#1#2{{\textstyle{#1\over\vphantom2\smash{\raise.20ex
        \hbox{$\scriptstyle{#2}$}}}}}                   
\def\sfrac#1#2{{\vphantom1\smash{\lower.5ex\hbox{\small$#1$}}\over
        \vphantom1\smash{\raise.4ex\hbox{\small$#2$}}}} 
\def\bfrac#1#2{{\vphantom1\smash{\lower.5ex\hbox{$#1$}}\over
        \vphantom1\smash{\raise.3ex\hbox{$#2$}}}}       
\def\afrac#1#2{{\vphantom1\smash{\lower.5ex\hbox{$#1$}}\over#2}}    

\def\[{\lfloor{\hskip 0.35pt}\!\!\!\lceil}
\def\]{\rfloor{\hskip 0.35pt}\!\!\!\rceil}

\def\du#1#2{_{#1}{}^{#2}}

\def\ha{{\fracmm12}}

\def\un{\underline}
\def\fracmm#1#2{{{#1}\over{#2}}}

\def\low#1{{\raise -3pt\hbox{${\hskip 0.75pt}\!_{#1}$}}}

\def\Dot#1{\buildrel{_{_{\hskip 0.01in}\bullet}}\over{#1}}

\newskip\humongous \humongous=0pt plus 1000pt minus 1000pt
\def\caja{\mathsurround=0pt}
\def\eqalign#1{\,\vcenter{\openup2\jot \caja
        \ialign{\strut \hfil$\displaystyle{##}$&$
        \displaystyle{{}##}$\hfil\crcr#1\crcr}}\,}
\newif\ifdtup

\def\plpl{\raise-2pt\hbox{$\raise3pt\hbox{$_+$}\hskip-6.67pt\raise0.0pt}}
\def\mimi{\raise-2pt\hbox{$\raise3pt\hbox{$_-$}\hskip-6.67pt\raise0.0pt}}

\def\dvm{\raisebox{-.45ex}{\rlap{$=$}}}
\def\DM{{\scriptsize{\dvm}}~~}
\def\lin{\vrule width0.5pt height5pt depth1pt}
\def\dpx{{{ =\hskip-3.75pt{\lin}}\hskip3.75pt }}

\def\ref#1{$\sp{#1)}$}

\def\pl#1#2#3{Phys.~Lett.~{\bf {#1}B} (19{#2}) #3}
\def\np#1#2#3{Nucl.~Phys.~{\bf B{#1}} (19{#2}) #3}

\def\cqg#1#2#3{Class.~and Quantum Grav.~{\bf {#1}} (19{#2}) #3}

\parskip=\medskipamount                 
\thicklines                         

\begin{document}


\thispagestyle{empty}               

\def\border{                                            
        \setlength{\unitlength}{1mm}
        \newcount\xco
        \newcount\yco
        \xco=-24
        \yco=12
        \begin{picture}(140,0)
        \put(-20,11){\tiny Institut f\"ur Theoretische Physik Universit\"at
Hannover~~ Institut f\"ur Theoretische Physik Universit\"at Hannover~~
Institut f\"ur Theoretische Physik Hannover}
        \put(-20,-241.5){\tiny Institut f\"ur Theoretische Physik Universit\"at
Hannover~~ Institut f\"ur Theoretische Physik Universit\"at Hannover~~
Institut f\"ur Theoretische Physik Hannover}
        \end{picture}
        \par\vskip-8mm}

\def\headpic{                                           
        \indent
        \setlength{\unitlength}{.8mm}
        \thinlines
        \par
        \begin{picture}(29,16)
        \put(75,16){\line(1,0){4}}
        \put(80,16){\line(1,0){4}}
      \put(85,16){\line(1,0){4}}
        \put(92,16){\line(1,0){4}}

        \put(85,0){\line(1,0){4}}
        \put(89,8){\line(1,0){3}}
        \put(92,0){\line(1,0){4}}

        \put(85,0){\line(0,1){16}}
        \put(96,0){\line(0,1){16}}
        \put(92,16){\line(1,0){4}}

        \put(85,0){\line(1,0){4}}
        \put(89,8){\line(1,0){3}}
        \put(92,0){\line(1,0){4}}

        \put(85,0){\line(0,1){16}}
        \put(96,0){\line(0,1){16}}
        \put(79,0){\line(0,1){16}}
        \put(80,0){\line(0,1){16}}
        \put(89,0){\line(0,1){16}}
        \put(92,0){\line(0,1){16}}
        \put(79,16){\oval(8,32)[bl]}
        \put(80,16){\oval(8,32)[br]}

        \end{picture}
        \par\vskip-6.5mm
        \thicklines}

\border\headpic {\hbox to\hsize{
\vbox{\noindent ITP--UH--01/96 \hfill February 1996 \\
hep-th/9602038 }}}

\noindent
\vskip1.3cm
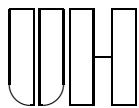
\begin{center}

{\Large\bf{ \noindent 2d, N=2~ AND ~N=4~ SUPERGRAVITY AND 
\vglue.1in
            THE LIOUVILLE THEORY IN SUPERSPACE~\footnote{
Supported in part by the `Deutsche Forschungsgemeinschaft',
the `Volkswagen Stiftung' and
\newline ${~~~~~}$ the NATO grant CRG 930789}}}\\
\vglue.3in

Sergei V. Ketov \footnote{
On leave of absence from:
High Current Electronics Institute of the Russian Academy of Sciences,
\newline ${~~~~~}$ Siberian Branch, Akademichesky~4, Tomsk 634055, Russia}

{\it Institut f\"ur Theoretische Physik, Universit\"at Hannover}\\
{\it Appelstra\ss{}e 2, 30167 Hannover, Germany}\\
{\sl ketov@itp.uni-hannover.de}
\end{center}

\vglue.3in

\begin{center}
{\Large\bf Abstract}
\end{center}

The two-dimensional (2d) manifestly locally supersymmetric actions describing the
N=2 and N=4 extended super-Liouville theory coupled to the N=2 and N=4 conformal 
supergravity, respectively, are constructed in superspace. It is shown that the 
N=4 super-Liouville multiplet is described by the {\it improved} twisted-II scalar
multiplet TM-II, whose kinetic terms are given by the $SU(2)\otimes U(1)$ WZNW 
model.

\newpage
\baselineskip=22pt

{\bf 1} {\it Introduction}. The non-critical N-extended superstrings are 
described by the 2d coupling of the N-extended supersymmetric scalar matter to 
the N-extended supergravity, with the exponential scalar potential. The standard
bosonic (N=0) Liouville action is given by
$$ I_0= \ha \int d^2x\sqrt{g} \left[ g^{\a\b}\pa_{\a}\f\pa_{\b}\f +QR^{(2)}\f
-\m^2 e^{\f}\right\}~.\eqno(1)$$
This action is invariant under the local Weyl transformations
$$ g_{\a\b}\to e^{2\s}g_{a\b}~,\qquad \f \to \f-2\s~,\eqno(2)$$
provided that the classical background charge $Q=2$, which is equivalent to 
demanding the metric $e^{\f}g_{\a\b}$ to be invariant.

It is rather straightforward to generalize the action (1) to the case of rigid 
or local (1,1) supersymmetry~\cite{n1gl,n1l}, but it is much less obvious for the
(2,2) and (4,4) {\it local} supersymmetry. The rigid manifestly supersymmetric 
actions describing the N=2 and N=4 super-Liouville theories are 
known~\cite{n2gl,n4gl} but, to the best of my knowledge, their locally 
supersymmetric counterparts were not constructed yet. Accordingly, the 
known studies of the N=2 and N=4 non-critical strings \cite{abk,kpr} were only 
performed in the superconformal gauge. It is the purpose of this letter to fill 
this gap. A manifestly covariant and supersymmetric formulation of the N=2 and 
N=4 supergravity, and of the related non-critical superstrings as well, can be 
useful for studies of super-Riemannian surfaces and super-Beltrami differentials,
and for calculating the non-critical superstring amplitudes, where the 
superconformal gauge is not convenient and a light-cone gauge is not accessible.

\vglue.2in

{\bf 2} {\it The N=2 Liouville Action}. The appropriate framework for describing
(2,2) supersymmetric matter couplings to (2,2) supergravity is provided by (2,2)
superspace. The necessary tools for that, such as superspace 
measures, invariant actions and component projection formulae, were recently 
developed by Grisaru and Wehlau \cite{gw}. 

The N=2 superspace has two real bosonic coordinates $x^{\dpx}$ and $x^{\DM}$,
and two complex fermionic coordinates $\q^+$ and $\q^-$, as well as their
conjugates $\q^{\Dot +}$ and $\q^{\Dot -}$. In addition to the (2,2) superspace
general coordinate transformations, the full local symmetries of the (nonminimal)
(2,2) supergravity include the local Lorentz symmetry, an axial $U_{\rm A}(1)$
 and a vector $U_{\rm V}(1)$ internal symmetries. The geometry of the (2,2) 
superfield supergravity is described in terms of the covariant spinorial 
derivatives
$$ \de_{\pm}= E_{\pm}{}^M\pa_M + \O_{\pm}\cm + \G_{\pm}\cx
+ \tilde{\G}_{\pm}\tilde{\cx}~, \eqno(3)$$
where the generators of the local Lorentz, $U_{\rm V}(1)$ and $U_{\rm A}(1)$
symmetries, $\cm$, $\cx$ and $\tilde{\cx}$, respectively, have been introduced.
The {\it minimal} $U_{\rm A}(1)$ version of the (2,2) supergravity is defined by 
the constraints ({\it cf.}\/ ref.~\cite{hp})
$$ \{ \de_{\pm }\, , \de_{\pm } \} ~=~ 0~,\quad
\{ \de\low{+ } \, , \de_{\Dot{+} } \} ~=~ i \de_{\dpx}~,\quad
\{ \de\low{- } \, , \de_{\Dot{-} } \} ~=~ i \de_{\DM}~,\quad
\{ \de_{+ } , \de_{\Dot -}  \} ~=~0~,$$
$$\{ \de_{+ } , \de_-  \} ~=~ - \frac{1}{2}R^*(\cm -i\cx)~,\eqno(4)$$
which leave a 2d `graviton', a complex `gravitino' and a $U(1)$ `graviphoton' as
the only component gauge fields in the theory. The (2,2) supergravity constraints
(4) allow the existence of the covariantly chiral superfields. It follows from 
the Bianchi identities that the superfield $R$ is covariantly chiral, 
$\de_{\Dot{\pm}}R=0$. The constraints (4) are also known to be invariant 
under the additional local Weyl (scale) transformations 
$$ E_{\pm}\to e^{\S} E_{\pm}~,\quad 
R^*\to e^{2\S}(R^* +  4 \[ \de_-,\de_+ \] \S)~,\eqno(5)$$
where the Weyl (2,2) superfield parameter $\S$ can be expressed in terms of a 
(2,2) chiral superfield~\cite{gw}.

The `component' complex curvature $R^{(2)}_{\rm c}$ of the (2,2) supergravity can
be most easily determined from the `space-time' commutator 
$$\eqalign{
 \[ \de_{\dpx},\de_{\DM} \] =~&~\ha\left(\de^2 R-\frac{1}{2}RR^*\right)(\cm+i\cx)
-\ha\left(\Bar{\de}^2R^* +\frac{1}{2} RR^*\right)(\cm -i\cx) \cr
~&~ +\ha(\de_+ R)\de_- + \ha(\de_- R)\de_+ -\ha(\de_{\Dot{+}}R^*)\de_{\Dot{-}}
-\ha( \de_{\Dot{-}}R^*) \de_{\Dot{+}}~,\cr} \eqno(6)$$
which is a consequence of the supergravity constraints (4). Therefore, we can
identify
$$\left(\de^2R-\frac{1}{2} RR^*\right)={\car}~,\quad  
\left.\car\right|=R^{(2)}_{\rm c}~,\eqno(7)$$
where $|$ means the projection onto the leading component of the superfield (we
ignore the gravitino contributions). The real part of the complex curvature 
$R^{(2)}_{\rm c}$ is just the usual 2d curvature $R^{(2)}$, whereas its imaginary
part is the abelian field strength of the graviphoton gauge field  from the 
(2,2) minimal supergravity multiplet. 

We are now in a position to write down the manifestly supersymmetric superfield 
action generalizing that of eq.~(1) to the case of (2,2) local supersymmetry. 
Since we are not interested in the most general matter couplings in (2,2) 
supergravity (see however, ref.~\cite{ggw}) but the Liouville-type interaction, 
we restrict ourselves to the (2,2) matter described by some chiral superfields 
$\f^a$. Their coupling to the (2,2) supergravity in (2,2) superspace is given by
$$ I_2= \int d^2x\left\{ \int d^2\q d^2\bar{\q}E^{-1}K(\f,\bar{\f}) -
\int d^2\q \ce^{-1}W(\f) - \int d^2\q \ce^{-1}R\U(\f) +{\rm h.c.}\right\}~,
\eqno(8)$$
where the supervielbein determinant $E(x,\q,\bar{\q})$, the chiral density 
$\ce(x,\q)$, the K\"ahler potential $K(\f,\bar{\f})$, the superpotential $W(\f)$, 
and the (holomorphic) dilaton $\U(\f)$ have been introduced. In components, after
using the projection formulae of ref.~\cite{gw}, the last term in eq.~(8) yields
$$\left.\int d^2x\int d^2\q \ce^{-1}R\U(\f)\right| =\left.\int d^2x e^{-1}
(\de^2-\frac{1}{2} R^*)R\U\right| = \eqno(9)$$
$$ = \left. \int d^2x e^{-1} \left( \car\U +\frac{1}{2}RR^*\U 
+R\de^2\U -\frac{1}{2} RR^*\U \right)\right| \equiv
\left.\int d^2x e^{-1} \left(\car\U + R\de^2\U\right)\right|~,$$
which proves that the non-propagating complex auxiliary field $H=\left.R\right|$ 
of the (2,2) supergravity multiplet enters the action (8) {\it linearly}. It is
also clear from eq.~(9) that the real part of the dilaton is coupled to the 2d
curvature, whereas its imaginary part is coupled to the abelian field strength of
the graviphoton. Evaluating the potential terms for the action (8) is easy, 
{\it viz}.
$$ 2\fracmm{\pa^2K}{\pa\f^a\pa\bar{\f}^b}F^aF^{*b} + \ha H^*W +
F^a\fracmm{\pa W}{\pa\f^a} + HF^a\fracmm{\pa\U}{\pa\f^a}+{\rm h.c.}~,
\eqno(10)$$
and it leads, after eliminating the matter complex auxiliary fields $F$, to the 
constraint 
$$ W=\left[ \fracmm{\pa^2 K}{\pa\f^a\pa\bar{\f}^b}\right]^{-1}
\fracmm{\pa\U^*}{\pa\bar{\f}^b}\left( \fracmm{\pa W}{\pa\f^a}
+H\fracmm{\pa\U}{\pa\f^a}\right)~,\eqno(11)$$
relating the superpotential with the dilaton, without using any gauge 
({\it cf}. ref.~\cite{abk}).  In the case of a {\it single} chiral 
superfield $\f$, it is always possible to make the dilaton field linear, $\U=\f$,
by field redefinition. Then eq.~(11) forces the K\"ahler metric to be flat, 
$K=\bar{\f}\f$, and implies $W(\f)=\m e^{\f}+H$, like in ref.~\cite{abk}. We can 
therefore conclude that the action (8) is the (2,2) locally supersymmetric 
generalization of eq.~(1) indeed.

\vglue.2in

{\bf 3} {\it The N=4 Liouville Action}. After the relatively simple N=2 exercise
given above, I turn to a construction of the N=4 generalization of the Liouville
action (1) in the curved superspace of 2d, (4,4) supergravity. 

The $N=4$ superspace is parametrized by the coordinates
$$ z^A ~=~ ( x^{\dpx}, x^{\DM}, \q^{+ i}, \q^{- i}, \q^{\Dot +}_{~~i},
        \q^{\Dot -}_{~~i} )~,\eqno(12) $$
where  $x^{\dpx}$ and $x^{\DM}$ are two real bosonic coordinates, $\q^{\pm i}$ 
and their complex conjugates $\q^{\Dot\pm}_{~~i}$ are complex fermionic 
coordinates, $i=1,2$. The fermionic coordinates $\q^{\pm i}$ are spinors with 
respect to $SU(2)$. Their complex conjugates are defined by
$$(\q^{\pm i})^* ~\equiv~\q^{\Dot \pm}_{~~i}~~~, \quad \q^{\Dot\pm ~i}=\cc^{ij}
\q^{\Dot \pm}_{~~j}~~~,\eqno(13)
$$
where the star denotes usual complex conjugation. The $SU(2)$ indices are raised 
and lowered by $\cc^{i j}$ and $\cc_{i j}$, whose explicit form is given by 
$\cc^{i j} = i\ve^{i j}$ and $(\cc^{i j})^* = \cc_{i j}$. 

The local symmetries of the minimal (4,4) superfield supergravity are the (4,4)
superspace general coordinate transformations, local Lorentz frame rotations and
$SU(2)$ internal frame rotations. Therefore, the fully covariant derivatives in
the curved (4,4) superspace should include the tangent space generators for all
that symmetries with the corresponding connections,
$$ \de_A ~=~ E_A^M D_M + \O_A \cm + i~ \G_{A}\cdot \cy~,\eqno(14)$$
where the (4,4) supervielbein $E_A^M$, the Lorentz generator $\cm$ with the
Lorentz connection $\O_A$, and the $SU(2)$ generators $\cy_i{}^j$ with the
$SU(2)$ connection  $(\G_{A})_j{}^i$ have been introduced. We use the notation
$\G_{A}\cdot\cy\equiv(\G_{A})_j{}^i\cy_i{}^j$.

Assuming that the supervielbein is invertible, the lowest-order
component in the $\q$-expansion of the superfield $E^a_{\m}$ can be identified 
with the zweibein, $\left. E^a_{\m}\right|=e^a_{\m}$. Similarly, 
$\left. E^{i\pm}_{\m}\right|=\j^{i\pm}_{\m}$ and 
$\left. \G\du{\m j}{i}\right|=B\du{\m j}{i}$ define the rest of the gauge fields
in the (4,4) supergravity multiplet. The superfield torsion and curvature
tensors are defined by
$$ \[ \de_A , \de_B \} = T\du{A B}{C}\de_C + R_{A B} \cm + i~ F_{A B}\cdot\cy~.
\eqno(15)$$

The generators for the local Lorentz and $SU(2)$ frame transformations can be 
defined, e.g., by their action on the spinorial derivatives,
$$ \[ \cm , \de\low{\pm i} \] ~=~ \pm \frac{1}{2} \de\low{\pm i}~,\quad
\[ \cm , \de_{\Dot{\pm}}^{~~i} \] ~=~ \pm \frac{1}{2} \de_{\Dot{\pm}}^{~~i}~,
$$
$$\eqalign{
\[ \cy_i^{~j} , \de\low{\pm k} \] ~=~& + \d_k^{~j} \de\low{\pm i} -
                      \frac{1}{2} \d_i^{~j} \de\low{\pm k} ~,\cr
\[ \cy_i^{~j} , \de_{\Dot{\pm}}^{~~k} ] ~=~&
      -  \d_i^{~k} \de_{\Dot{\pm}}^{~~j} +
                      \frac{1}{2} \d_i^{~j} \de_{\Dot{\pm}}^{~~k} ~.\cr}
\eqno(16)$$

The superspace constraints defining the (4,4) minimal supergravity were first
formulated by Gates {\it et. al.} in ref.~\cite{glo}. In our notation, they take 
the form
$$
\{ \de_{\pm i}\, , \de_{\pm j} \} ~=~ 0~,\quad
\{ \de\low{+ i} \, , \de_{\Dot{+} j} \} ~=~ i \cc\low{i j} \de_{\dpx}~,\quad
\{ \de\low{- i} \, , \de_{\Dot{-} j} \} ~=~ i \cc\low{i j} \de_{\DM}~,$$
$$\{ \de_{+ i} \, , \de_-^{~~j}  \} ~=~ - \frac{i}{2}~R^*~
       \left( \d_i^{~j}\cm - \cy_i^{~j} \right) ~,$$
$$ \{ \de\low{+ i}\, , \de_{\Dot -}^{~~j}  \} ~=~
      - \frac{i}{2}~ S~ \left( \d_i^{~j}\cm - \cy_i^{~j} \right) -
    \frac{1}{2}~ T~ \left( \d_i^{~j}\cm - \cy_i^{~j} \right) ~,\eqno(17)$$
plus their complex conjugates, where the four scalar (4,4) covariant field 
strength superfields have been introduced, the complex one, $R$, and the two real
ones, $S$ and $T$. Using the Bianchi identities associated with the constraints 
(17), one finds that those superfields satisfy the constraints~\cite{glo,sm} 
$$
\de_{\Dot{\pm} i} {R} ~=~0~,\quad
\de\low{{\pm} i}  {R} ~=~ \pm ~ \de_{\Dot{\pm} i} S~,\quad
\de_{{\pm} i} S ~=~ \pm ~i~ \de_{{\pm} i} T~,\eqno(18)
$$
which are the defining constraints of the 2d, (4,4) {\it twisted}-I off-shell
hypermultiplet (TM-I) according to the classification of ref.~\cite{gk4}. The
independent components of the TM-II are determined by the constraints (18) and
they comprise, in addition to the leading scalars $(R,R^*,S,T)$, a complex spinor
$SU(2)$ doublet $\j^i_{\pm}$, a real auxiliary singlet $A$ and an auxiliary 
triplet $A\du{i}{j}$. The supersymmetry transformation laws of the TM-I 
components can be found, e.g., in ref.~\cite{gk4}. This 2d, (4,4) off-shell 
hypermultiplet can be obtained via dimensional reduction from the 4d, N=2 
off-shell vector multiplet~\cite{ga}.  

As far as the `component' quaternionic curvature of the (4,4) supergravity is 
concerned, it can be deduced from the `space-time' commutator~\footnote{The full
superspace structure of 2d, (4,4) supergravity will be considered in a separate
publication~\cite{sm}.} 
\begin{eqnarray}
\[ \de_{\dpx} , \de_{\DM} \] &=&
\fracmm{i}{4}~ \Bigl[ ~
- ( \de_{-}^{~~i}{R} ) \de_{+i}
+ ( \de_{+}^{~~i}{R} ) \de_{-i}
+ ( \de_{\Dot-}^{~~i} R^* ) \de_{\Dot+ i}
- ( \de_{\Dot+}^{~~i} R^* ) \de_{\Dot- i}  \nonumber \\
&&~~~~~~~~
+ ( \de_{\Dot-}^{~~i}~ (S+iT) )  \de_{+i}
- ( \de_{-}^{~~i}~ (S- iT) )  \de_{\Dot+ i} \nonumber \\
&&~~~~~~~~
+ ( \de_{\Dot+}^{~~i} ~ (S- iT) )  \de_{-i}
- ( \de_{+}^{~~i}~ (S+iT) )  \de_{\Dot- i}
~\Bigr]         \nonumber \\
&&
+\fracmm{1}{2}~ \Bigl[ ~
  R R^* - S^2 - T^2
+\frac{i}{4}  ( \de_{+}^{~~i} \de_{-i} R )
-\frac{i}{4}  ( \de_{\Dot+}^{~~i} \de_{\Dot- i} R^* )  \nonumber \\
&&~~~~~~~~
+\frac{i}{4}  ( \de_{\Dot+}^{~~i} \de_{-i}~ (S- iT) )
-\frac{i}{4}  ( \de_{+}^{~~i} \de_{\Dot- i}~  (S+iT) )
~\Bigr] ~  \cm   \nonumber \\
&&
+\fracmm{i}{8}   \Bigl[ ~
 ( \de_{+}^{~~i} \de_{-j} R )
-  ( \de_{\Dot+}^{~~i} \de_{\Dot- j} R^* )  \nonumber \\
&&~~~~~~~~
+ ( \de_{\Dot+}^{~~i} \de_{-j}~ (S- iT) )
-  ( \de_{+}^{~~i} \de_{\Dot- j}~  (S+iT) )
~\Bigr] ~\cy_i^{~j}
 ~ ,   \nonumber
\end{eqnarray}
$$
\[ \de_{\dpx} , \de_{\dpx} \] ~=~ 0~,\qquad
\[ \de_{\DM} , \de_{\DM} \] ~=~ 0~. \eqno(19)$$
The real part of the quaternionic curvature (in front of $\cm$) contains 
$R^{(2)}$ in its leading component, whereas the imaginary part (in front of the
$SU(2)$ generator $\cy\du{i}{j}$) has the $SU(2)$ field strength $F\du{j}{i}(B)$ 
in its leading component.

There exist two known $(8+8)$ off-shell versions of 2d, (4,4) hypermultiplet 
(with finite number of auxiliary fields), TM-I and TM-II. The TM-I was already 
introduced above. The {\it twisted}-II off-shell hypermultiplet (TM-II) in 2d was
discovered by Ivanov and Krivonos~\cite{n4gl}. The covariant superspace  
constraints defining this version of (4,4) hypermultiplet are given by
$$
\de\low{\pm i}~ L\du{j}{k} = \mp i \left( \d\du{i}{k} \de\low{\pm j} P 
- \frac{1}{2} \d\du{j}{k}  \de\low{\pm i} P  \right) ~,\eqno(20)
$$
in terms of 1+3 scalar superfields $P$ and $L\du{i}{j}$, where $P^*=P$,  
$L\du{i}{j}=(L\du{j}{i})^*$ and $L\du{i}{i}=0$. It follows from eq.~(20) that the
independent components of TM-II comprise, in addition to the  leading components 
$P$ and $L\du{i}{j}$, a complex spinor doublet $\l^i_{\pm}$, and four auxiliary
scalars: a complex one $G$ and two real ones $M$ and $N$. The supersymmetry 
transformation laws of the TM-II components can be found e.g., in ref.~\cite{gk4}.
 The TM-II can be obtained via dimensional reduction from the 4d, N=2 off-shell 
tensor multiplet.

Among the two hypermultiplets TM-I and TM-II, only the latter can represent the 
(4,4) Liouville multiplet because of the $SU(2)$ structure of their leading 
components. It becomes obvious by noticing that the (4,4) supergravity constraints
(17) are invariant under the (4,4) super-Weyl transformations having the 
form~\cite{ghn,sm}
$$ \de\low{\pm i} \rightarrow  \frac{1}{2} P~ \de\low{\pm i}
\mp i L\du{i}{j} ~\de\low{\pm j}  \mp ( \de\low{\pm i} P ) \cm +
         ( \de\low{\pm j} P ) \cy\du{i}{j}~,\eqno(21) $$
where the super-Weyl infinitesimal (4,4) superfield parameters $(P,L\du{i}{j})$ 
form a TM-II multiplet.

The auxiliary fields of TM-II can be considered as the leading components of 
a TM-I which can be called the kinetic multiplet, like in 4d. Therefore, TM-I and
TM-II are dual to each other, though they are not equivalent~\cite{gk4}. It 
implies, in particular, that there exists the supersymmetric invariant given by 
a product of TM-I and TM-II, without the introduction of a central 
charge~\cite{n4gl}. In a curved superspace, this invariant takes the form
$$ I = \int d^2x d^4\q d^4\bar{\q}E^{-1} (\P S + \X T) +\left[  \int d^2x d^4\q 
\ce^{-1} \L R + {\rm h.c.}\right]~,\eqno(22)$$
where the (4,4) supervielbein determinant $E^{-1}$, the chiral super-density 
$\ce^{-1}$, the real superfield prepotentials $\P$ and $\X$, and the chiral
superfield prepotential $\L$ of the TM-II have been introduced~\cite{gk4}.

The rigidly (4,4) supersymmetric invariant describing the free TM-II action, 
which is quadratic in the fields, is known~\cite{gk4}. However, its locally 
(4,4) supersymmetric generalization does not exist.~\footnote{The same is true in
4d.} The allowed matter couplings in the (4,4) conformal supergravity are much 
more restricted than in the rigid (4,4) case, and it is also known to be the case
for the N=2 matter couplings in 4d, N=2 supergravity.~\footnote{See, e.g., 
ref.~\cite{vanp} for a recent review.} Fortunately, as far as the TM-II in 2d is
concerned, this problem is only apparent, since there exists its {\it improved} 
2d version, which can be coupled to the 2d, (4,4) conformal supergravity~! 
Namely, it is possible to form the TM-I out of the TM-II components in yet 
another {\it non-linear} way:
$$
\eqalign{  
R_{\rm impr.}~=~&~ L^{-1} \left( G -2L\du{i}{j} A\du{j}{i} \right) 
-2i \l\low{i\Dot{+}} {\l^j}\low{\Dot{-}} L\du{j}{i} L^{-3}~,\cr
S_{\rm impr.}~=~&~ L^{-1} \left( M - 4PA \right)
-i \left( \l\low{i\Dot{+}}{\l^j}\low{-} - \l\low{i\Dot{-}} {\l^j}\low{+}\right)
L\du{j}{i} L^{-3}~,\cr  
T_{\rm impr.}~=~&~ L^{-1}\left( N - 4PA \right)
+ \left( \l\low{i\Dot{+}}{\l^j}\low{-} + \l\low{i\Dot{-}} {\l^j}\low{+} \right)
  L\du{j}{i} L^{-3}~,\cr}\eqno(23)$$
where 
$$ L\equiv \sqrt{L\du{i}{j}(L\du{i}{j})^*}~.\eqno(24)$$
Eq.~(22) can now be used to define the invariant coupling of the improved TM-II 
to the (4,4) supergravity, in the form
$$ I_{\rm impr.} = \int d^2x d^4\q d^4\bar{\q}E^{-1} (\P S_{\rm impr.} 
+ \X T_{\rm impr.}) +\left[  \int d^2x d^4\q\ce^{-1} 
\L R_{\rm impr.} + {\rm h.c.}\right]~.\eqno(25)$$

The existence of the improved TM-II in 2d is a direct consequence of the existence
of the improved N=2 tensor multiplet in 4d~\cite{wpp}, since they are related via
dimensional reduction. Unlike the improved N=2 tensor multiplet in 4d, its 2d
counterpart does not have any gauge degrees of freedom, which allows the (4,4)
locally supersymmetric component action associated with eq.~(25) to have manifest
$SU(2)$ internal symmetry. 

The action (25) is highly non-linear even in the absence of supergravity. The 
kinetic terms of the improved TM-II in eq.~(25) are given by the non-linear 
sigma-model (NLSM) {\it with} torsion. A direct calculation yields
$$L_{NLSM} = -\ha g_{ab}\pa_{\a}X_a\pa^{\a}X_b -\ha i\ve^{\a\b}e_{ab}
\pa_{\a}X_a\pa_{\b}X_b~,\eqno(26)$$
where the symmetric metric $g_{ab}(X)$ and antisymmetric torsion potential 
$e_{ab}(X)$ are given by
$$ g_{ab} =e^{-\f}\d_{ab}~,\qquad e_{IJ}=-X_4\ve_{IJK}X_Ke^{-3\f}~,
\quad e_{I4}=0~,\eqno(27)$$
when using the notation $a,b=I,4$, and $I,\ldots=1,2,3$, and
$$L_{ij} =\left( \begin{array}{cc} X_1 + iX_2 & iX_3 \\
iX_3 & X_1 -iX_2 \end{array} \right)~,\quad P=X_4~,\quad 
L=\sqrt{X_1^2+X_2^2+X_3^2}\equiv e^{\f}~.\eqno(28)$$

The NLSM of eq.~(26) is quite remarkable. First, its four-dimensional
target space has the explicit $SO(3)\simeq SU(2)$ internal symmetry. Second, the
target space metric is conformally flat, and it does not depend on $X_4$. Third, 
despite of the apparent dependence of the torsion potential $e_{ab}$ upon all of 
the coordinates, the geometrical torsion $H_{abc}=\frac{3}{2}\pa_{[a}e_{bc]}$ 
does {\it not} actually depend upon $X_4$, and, hence, the NLSM target space has 
an abelian isometry. Third and the most importantly, the torsion $H_{abc}$ is 
just the {\it parallelizing} torsion for the curvature associated with the metric,
 i.e. the generalized Riemann curvature tensor (with torsion) identically zero in
this case. Since the group manifolds are the only parallelizable manifolds in 
four dimensions, the symmetry considerations dictate that eq.~(26) is just the 2d 
Wess-Zumino-Novikov-Witten (WZNW) model over the $SU(2)\otimes U(1)$ in the rather
unusual parametrization of the target space~!

We are now in a position to construct the manifestly locally (4,4) supersymmetric 
Liouville action. The quaternionic curvature of the (4,4) supergravity enters the
action (25) linearly, being multiplied by the physical scalar components of the
improved TM-II modulo (4,4) super-Weyl rescaling of the supergravity fields. 
Therefore, we only need to find a proper term which would generate the 
Liouville-type potential. As is well-known, there is, in fact, the additional 
natural resource to build a supersymmetric invariant, namely, the 
Fayet-Iliopoulos (FI) term.~\footnote{This term was also used in ref.~\cite{n4gl}
to construct the rigidly (4,4) supersymmetric Liouville action.} In (4,4) 
superspace, the FI term takes the form 
$$ I_{\rm FI} = -2\m  \int d^2x d^4\q d^4\bar{\q}E^{-1} \P~.\eqno(29)$$
Given the action
$$ I_4=I_{\rm impr.} + I_{\rm FI}~,\eqno(30)$$
the auxiliary field $P$ of the improved TM-II enters this action in the 
combination $e^{-\f}P^2-2\m P$. Eliminating the auxiliary field $P$ via its 
algebraic equation of motion just yields the desired contribution $-\m^2e^{\f}$. 
We can therefore conclude that the action (30) is the (4,4) locally 
supersymmetric Liouville action indeed.

\vglue.2in

{\bf 4} {\it Conclusion}. It is well-known that the $SU(2)\otimes U(1)$ WZNW model
has the classical (4,4) superconformal invariance, which is represented by the
`large' N=4 superconformal algebra in the left- or right-moving sector~\cite{stp}.
The (4,4) conformal supergravity respects only the `small' $SU(2)$ part of this 
N=4 algebra. The fact that the kinetic terms of the improved TM-II action are 
given by the WZNW model allows one to apply the methods of conformal field 
theory~\footnote{See ref.~\cite{book} for a recent review, or an introduction.}
for an investigation of the quantum (4,4) Liouville theory. In ref.~\cite{kpr}, 
this fact was not proved but used as the basic assumption. 

For a detailed analysis of the (4,4) Liouville theory in a curved superspace of
the (4,4) supergravity, one needs to develop further the superspace formalism,
e.g. to solve the superspace constraints and to calculate the component projection
formulae, which we postpone for another publication~\cite{sm}. The related quantum
aspects of the N=4 Liouville theory also deserve further studies~\cite{su}.
\vglue.2in
 
{\it Acknowledgment}. The author would like to thank the Physics Departments of
Brandeis and Maryland Universities, where this work was initiated, for 
hospitality, and thank Jim Gates, Marc Grisaru, Olaf Lechtenfeld and Sven Moch
for discussions.
\vglue.2in

\end{document}